\documentclass[nofootinbib,prd, preprint]{revtex4-1}
\usepackage{graphicx}
\usepackage{longtable}
\usepackage{amsmath}
\usepackage{amssymb}
\usepackage{subfigure}
\usepackage{hyperref}
\usepackage{stackengine}
\usepackage{color}

\newcommand{\mpl}{{M_{\rm {pl}}}}

\newcommand{\dd}{\, {\rm d}}

\newcommand{\pn}{\Phi_{\rm N}}

\newcommand{\oo}{\mathcal{O}}

\newcommand{\eff}{_{\rm eff}}

\newcommand{\oxx}{\Omega_X}

\newcommand{\oscc}{^{\rm osc}}
\newcommand{\e}{\varepsilon}

\newcommand{\osc}{_{\rm osc}}
\newcommand{\nbl}{n_{B-L}}
\newcommand{\Om}{\Omega}
\newcommand{\rve}{\right\vert}
\newcommand{\lve}{\left\vert}

\newcommand{\thee}{\theta_\varepsilon}

\begin{document}
\title{Baryogenesis via Dark Matter-Induced Symmetry Breaking in the Early Universe}
\author{Jeremy Sakstein}
\email[Email: ]{sakstein@physics.upenn.edu}
\affiliation{Center for Particle Cosmology, Department of Physics and Astronomy, University of Pennsylvania 209 S. 33rd St., Philadelphia, PA 19104, USA}
\author{Mark Trodden}
\email[Email: ]{trodden@physics.upenn.edu}
\affiliation{Center for Particle Cosmology, Department of Physics and Astronomy, University of Pennsylvania 209 S. 33rd St., Philadelphia, PA 19104, USA}

\begin{abstract}
We put forward a new proposal for generating the baryon asymmetry of the universe by making use of the dynamics of a $\mathrm{U}(1)$ scalar field coupled to dark matter. High dark matter densities cause the $\mathrm{U}(1)$ symmetry to break spontaneously so that the field acquires a large vacuum expectation value. The symmetry is restored when the density redshifts below a critical value, resulting in the coherent oscillation of the scalar field. A net $B-L$ number can be generated either via baryon number-conserving couplings to the standard model or through small symmetry-violating operators and the subsequent decay of the scalar condensate.\end{abstract}

\maketitle

\section{Introduction} 
The origin of the matter-antimatter asymmetry that we observe today remains a mystery. That it is simply an initial condition seems unlikely since inflation would have diluted any initial asymmetry to negligible values, and so a dynamical generation mechanism---baryogenesis (for reviews see, e.g.~\cite{Trodden:1998ym,Riotto:1999yt,Dine:2003ax,Cline:2006ts,Morrissey:2012db,Allahverdi:2012ju})---operating sometime between reheating and the electroweak phase transition\footnote{Although mechanisms that operate at lower temperatures do exist ~\cite{Krauss:1999ng,GarciaBellido:1999sv,Copeland:2001qw}.} seems unavoidable. Electroweak baryogenesis in the standard model cannot account for the large asymmetry observed today (see \cite{Morrissey:2012db} and references therein), and so one is naturally led to some combination of beyond the standard model physics and exotic asymmetry generation mechanisms. 

Some of these rely on a breaking of $\mathrm{U}(1)_{B-L}$ in the early Universe and its subsequent restoration at later times. Such a breaking has hitherto been achieved using either a coupling of a fundamental $\mathrm{U}(1)$ scalar $\phi$ to the inflaton \cite{Affleck:1984fy} or through thermal corrections \cite{Dodelson:1989ii} to the potential. In this paper, we propose an alternative mechanism. A Weyl coupling of a $\mathrm{U}(1)$ charged scalar $\phi$ (with bare mass $m_\phi$) to dark matter so that the effective metric for dark matter is 
\begin{equation}\label{eq:Weyl}
\tilde{g}_{\mu\nu}=A^2(|\phi|)g_{\mu\nu};\quad A(|\phi|)=1-\frac{|\phi|^2}{3M^2},
\end{equation}
results in an effective mass for the scalar
\begin{equation}\label{eq:newmass}
m\eff^2=m_\phi^2-\frac{\rho_{\rm DM}}{3M^2}.
\end{equation}
Consequently, the symmetry is broken at high dark matter densities (when $\rho_{\rm DM}\gg m_\phi^2M^2$) and restored as the dark matter naturally redshifts. In what follows, we will illustrate how such a symmetry breaking scenario can give rise to baryogenesis using a common and well-studied paradigm: the Affleck-Dine (AD) mechanism \cite{Affleck:1984fy}.

In the AD mechanism, the coherent oscillation of a scalar field charged under $\mathrm{U}(1)_{B-L}$ about its symmetry restoring minimum can generate a net $B-L$ charge through symmetry violating terms. In the context of the original model, the scalar represents a flat direction in the minimal supersymmetric standard model (MSSM) and symmetry violation is present due to non-renormalisable terms in the superpotential that arise from supersymmetry breaking \cite{Dine:2003ax,Dine:1995kz}. The initial condition is provided by a breaking of the symmetry during inflation. In particular, a coupling of the scalar to the inflaton via the K\"{a}hler potential gives rise to an effective mass term for the scalar
\begin{equation}\label{eq:ADmass}
m\eff^2=m_{3/2}^2-c\frac{\rho_I}{\mpl^2} \ ,
\end{equation}
where $m_{3/2}$ is the gravitino mass (this term arises from supersymmetry breaking), $\rho_I$ is the energy density of the inflaton, and $c\sim\oo(1)$. At the end of inflation the inflaton undergoes coherent oscillations about the minimum of its potential and its energy density redshifts like matter. Initially, one has $\rho_I\gg m_{3/2}^2\mpl^2$ and the effective mass is tachyonic so that the field tracks a time-dependent symmetry breaking minimum. When the inflaton has redshifted such that $\rho_I\approx m_{3/2}^2\mpl^2$, the mass is positive and the symmetry is restored, spurring the field into coherent oscillations about the symmetry restoring minimum at zero field value.

The scenario we describe in this paper can be thought of as new and general mechanism for decoupling some of the features of Affleck-Dine baryogenesis from the details of inflation. The mass of the scalar field in \eqref{eq:newmass} has a similar form to the mass of the AD field in \eqref{eq:ADmass}, allowing for the symmetry-breaking mechanism we employ to play an important role in baryogenesis. This may allow for more complicated reheating scenarios such as preheating. Furthermore, the phenomenology of the resulting model is different from the AD case, since the density that breaks the symmetry need not dominate the energy budget of the Universe, allowing for baryogenesis to take place during the radiation-dominated epoch rather than during perturbative reheating, during which the expansion rate takes a brief, matter-dominated form. 

In the remainder of this paper we will illustrate how our mechanism may be implemented into an Affleck-Dine type model. We will use a simple toy model to investigate the new phenomenology, and to illustrate the new features that arise. We calculate important quantities such as the baryon to photon ratio, and discuss possible ways that the baryon asymmetry can be transferred to the visible sector. Our goal here is not to concoct a new cosmological model, but rather to demonstrate the mechanism in a familiar and simple scenario. In particular, we note that there are many other choices one \emph{could} make instead, and many assumptions that one \emph{could} drop in favor of others.  In each case, one should be able to construct alternative viable scenarios. We end by discussing possible tests of our new mechanism, and how one could implement it into other baryogenesis scenarios, in particular, baryon-symmetric baryogenesis \cite{Dodelson:1989ii}.

\section{Toy Model} 
Our simple toy model is
\begin{equation}\label{eq:toy}
S=\int\dd^4x\sqrt{-g}\left(\mathcal{L}_{\rm GR} + \mathcal{L}_\phi+\mathcal{L}_X+\mathcal{L}_{X\phi}\right),
\end{equation}
where $\mathcal{L}_{\rm GR}$ is the Einstein-Hilbert action, and
\begin{equation}
\mathcal{L}_\phi = -\partial_\mu\phi\partial^\mu\phi^\dagger - m_\phi^2\lve\phi\rve^2-\lambda\frac{\lve\phi\rve^{2n}}{\mu^{2n-4}}-\frac{\varepsilon}{4}\phi^4-\frac{\varepsilon^\dagger}{4}{\phi^\dagger}^4,
\end{equation}
is the Lagrangian for a complex scalar field charged under a $\mathrm{U}(1)$ symmetry, which we take to be $\mathrm{U}(1)_{B-L}$, with small symmetry breaking terms proportional to $\varepsilon$. Here, $\mathcal{L}_X$ is the Lagrangian for a Dirac fermion $X$ that plays the role of dark matter, which has mass $m_X$, and 
\begin{equation}\label{eq:Weylferm}
\mathcal{L}_{X\phi}= m_X\frac{|\phi|^2}{3M^2}\bar{X}X
\end{equation}
is an interaction between dark matter, mediated by the scalar field, that is equivalent to a coupling of dark matter to the metric \eqref{eq:Weyl} \cite{Wetterich:2014bma}\footnote{The nature of dark matter is unimportant for the success of the mechanism; all that is required is a tachyonic mass for the scalar proportional to the dark matter density, which can always be achieved through a Weyl coupling to dark matter given by equation \eqref{eq:Weyl}. We choose a Dirac fermion here for the purposes of providing a concrete model.}. The dynamics of the scalar are governed by the effective potential
\begin{equation}\label{eq:Veff}
V(\phi,\phi^\dagger)\eff= \left(m_\phi^2-\frac{\rho_X}{3M^2}\right)|\phi|^2 +\lambda\frac{\lve\phi\rve^{2n}}{\mu^{2n-4}}+\frac{\varepsilon}{4}\phi^4+\frac{\varepsilon^\dagger}{4}{\phi^\dagger}^4.
\end{equation}
Note that when $n=2$ the potential is renormalisable $\lambda|\phi|^4$, whereas when $n>2$ we can, and will, set $\lambda=1$ without loss of generality. In the standard Affleck-Dine scenario, the case $n=2$ is typically not viable because, in a matter-dominated era, the field does not track its minimum prior to symmetry restoration \cite{Dine:1995kz}. However, our alternative scenario takes place during a radiation-dominated era in which the minimum of the effective potential is an attractor. We will verify this numerically later.

The quintessential features of the mechanism are now clear: when $\rho_X\gg M^2\mpl^2$ the field rolls to the minimum of the effective potential given by (ignoring the small symmetry-violating terms proportional to $\varepsilon$)
\begin{equation}\label{eq:phimin1}
|\phi|_{\rm min}=\begin{cases}
    \sqrt{\frac{\rho_X}{6\lambda M^2}}       & \quad n=2\\
    \left(\frac{\rho_X\mu^{2n-4}}{3nM^2}\right)^{\frac{1}{2n-2}}  & \quad n>2\\
  \end{cases} \ ,
\end{equation}
where the $\mathrm{U}(1)_{B-L}$ is broken. As the dark matter redshifts, the symmetry is restored and the field now rolls towards the new minimum at $|\phi|=0$. This mechanism is reminiscent of the symmetron mechanism~ \cite{Hinterbichler:2010es,Hinterbichler:2011ca}, often studied in the context of modified gravity, where a Weyl coupling similar to \eqref{eq:Weyl} is used to implement a density-dependent breaking of a $\mathbb{Z}_2$ symmetry of a real scalar. However, in the present case the situation is reversed, so that the symmetry is restored at high densities. In this sense, our model is more reminiscent of the asymmetron model~\cite{Chen:2015zmx}. The effective potential \eqref{eq:Veff} is strictly a toy model that we will use to illustrate the mechanism, which can be implemented in any model in which the symmetry breaking is density-dependent. In particular, one could consider the $\mathrm{U}(1)$ generalizations of radiatively stable symmetrons \cite{Burrage:2016xzz} or generalized symmetrons \cite{Brax:2012gr}, where the symmetry breaking is due to higher-order operators. 

\section{Cosmology and Baryogenesis} \label{sec:cosmo}
The simplest scenario, which we will consider here, is that the products of reheating/preheating are radiation, which dominates the Universe, a small component of dark matter $X$, and the scalar $\phi$. Setting $\rho_X=3\Omega_X\mpl^2H^2$, the symmetry-breaking minimum \eqref{eq:Veff} corresponds to
\begin{equation}\label{eq:phimin2}
|\phi|_{\rm min}=\begin{cases}
    \alpha H\sqrt{\frac{\Omega_X}{2\lambda}}       & \quad n=2\\
   \mu\left(\frac{\alpha^2\Om_XH^2}{n\mu^2}\right)^{\frac{1}{2n-2}}  & \quad n>2\\
  \end{cases},
\end{equation}
where $\alpha\equiv\mpl/M$. Note that since we are in the radiation-dominated era we have $\Omega_X\ll1$ and so the Hubble friction $H\dot{\phi}\sim H^2\phi$ will result in the field being over-damped and remaining at its initial value unless the driving term $\Omega_X\alpha^2H^2$ dominates, in which case the field will roll to track the minimum. This requires $\alpha\gg1$ and so, in particular, gravitational strength couplings $\alpha=1$ ($M\sim \mpl$) cannot give rise to any appreciable asymmetry. This is in contrast to the original Affleck-Dine scenario, where the coupling is gravitational but the driving term can be large because the field is coupled to the particle that dominates the evolution of the Universe (the inflaton). 

The symmetry is restored when $H=H\osc$ with
\begin{equation}
\sqrt{\Omega_X}H\osc=\frac{m_\phi}{\alpha},
\end{equation}
and we denote the value of $\Omega_X$ when this happens by $\Omega^{\rm osc}_X$. Writing $\phi=|\phi|e^{i\theta}$ and $\varepsilon=\varepsilon_0e^{i\thee}$, the symmetry-violating terms introduce a correction to the potential given by 
\begin{equation}\label{eq:Vcorr}
\Delta V = \frac{\varepsilon_0|\phi|^4}{2}\cos(4\theta+\thee) \ .
\end{equation}
We require these terms to become important at $H=H\osc$, in order to generate the asymmetry. This happens when\footnote{Note that $\partial_\mu\phi\partial^\mu\phi^\dagger\supset |\phi|^2(\partial_\mu\theta)^2$ so that one must canonically normalise the field with a factor of $|\phi|^{-2}$.} $m_\theta^2=\Delta V''(\theta)/|\phi|^2\sim\varepsilon_0|\phi|^2$ is of order $H\osc^2$ and so we require
\begin{equation}\label{eq:epssim}
\varepsilon_0\sim\frac{1}{\alpha^2\Omega_X\oscc}\begin{cases}
    \lambda    & \quad n=2\\
   \left(\frac{m_\phi}{\mu}\right)^{\frac{2n-4}{n-1}} & \quad n>2\\
  \end{cases}.
\end{equation}

If the field has charge $q$ under $\mathrm{U}(1)_{B-L}$ then the conserved charge density is $\nbl=J^0=iq(\phi^\dagger\overset{\leftrightarrow}{\partial^0}\phi)=2q|\phi|^2\dot{\theta}$ and, using the equation of motion for the angular field,
\begin{equation}
|\phi|^2\left(\ddot{\theta}+3H\dot{\theta}\right)+2\chi\dot{\theta}=\varepsilon_0|\phi|^4\sin(4\theta+\thee),
\end{equation}
we find
\begin{equation}\label{eq:neqn}
\dot{n}_{B-L}+3H\nbl=2q\varepsilon_0|\phi|^4\sin(4\theta+\thee).
\end{equation}
At this point, in order to obtain an analytic solution, and to gain some insight into how the mechanism operates, we make the approximation that $\dot{n}_{B-L}\approx H\nbl$ \cite{Asaka:2000nb,vonHarling:2012yn}. We will return later in the paper to check numerically that the results we obtain by using this approximation can be trusted. Making our approximation, we obtain
\begin{equation}
\nbl=\frac{\e_0q|\phi|^4}{H\osc}\sin(4\theta_0+\theta_\e).
\end{equation}
Using equations \eqref{eq:phimin2} and \eqref{eq:epssim} we then have
\begin{equation}\label{eq:nbl}
\nbl \sim \begin{cases}
    \frac{m_\phi^3}{\lambda \alpha\sqrt{\oxx\oscc}}   & \quad n=2\\
   \frac{\mu^3}{\alpha\sqrt{\Omega_X\oscc}}\left(\frac{m_\phi}{\mu}\right)^{\frac{n+1}{n-1}} & \quad n>2\\
  \end{cases},
\end{equation}
where we have assumed $q\sim\oo(1)$. Ultimately, this will be transferred to the visible sector (the details of which we discuss later) and converted to baryon number $n_B$ via Sphaleron processes \cite{Harvey:1990qw} so that $n_{B}\sim\nbl$. 

It is worth noting that  at this point we could determine $\lambda$ or $\mu$ by demanding that $\Omega_B/\Omega_{\rm DM}$ has the correct value that we observe today. In order to emphasize the generality of the mechanism, we will not make this choice here. The dark matter density today is highly model-dependent and depends on physics that is not necessarily connected to baryogenesis, such as freeze-out. Furthermore, it is not even necessary for $X$ to be the particle that comprises the majority of the dark matter dominating our Universe today. It could instead simply be an inflaton decay product that itself decays, or that ends up being sub-dominant to other particles produced later on, possibly through the decay of the scalar condensate. For these reasons, we prefer not to commit to a specific scenario and restrict our model unnecessarily.

If we assume that the Universe reheats instantaneously at a temperature $T_R$ when $H=H_R\sim H\osc$\footnote{The latter assumption is not strictly necessary but it allows for a simple calculation of the baryon to photon ratio that is not tied to the details of the cosmic expansion.}, then the entropy density is %
\begin{equation}
s = \frac{4\rho_I}{3T_R} \sim \frac{H\osc^2\mpl^2}{T_R}\ ,
\end{equation}
in which case we can use equation \eqref{eq:nbl} to find
\begin{align}\label{eq:nblfinal}
\frac{\nbl}{s}\sim &10^{-10}\alpha\sqrt{\oxx\oscc}\frac{T_R}{10^8\textrm{ GeV}}\begin{cases}
  \frac{m_\phi}{\lambda\mpl}  & \quad n=2\\
   \frac{\mu}{\mpl}\left(\frac{\mu}{m_\phi}\right)^{\frac{n-3}{n-1}} & \quad n>2\\
  \end{cases}.
\end{align}
One can see that we can reproduce the observed baryon to photon ratio with a suitable choice of $T_R$, $\oxx\oscc$, and $\lambda$ or $\mu$. Note that imposing the condition $H_R\sim H\osc$ gives us the condition
\begin{equation}\label{eq:mphi}
\frac{m_\phi}{0.1\textrm{ GeV}}\sim \frac{\alpha}{10^{12}}\sqrt{\frac{\oxx\oscc}{10^{-18}}}\left(\frac{T_R}{10^{8}\textrm{ GeV}}\right)^2,
\end{equation}
where we have used the fact that $H^2\approx\pi^2 T^4/45\mpl^2$ in a radiation dominated Universe. We also have
\begin{equation}\label{eq:check}
\Omega_\phi\approx\frac{m_\phi^2|\phi|^2}{3H\osc^2\mpl^2}\sim \oxx\frac{|\phi|^2}{M^2}\ll1 \ ,
\end{equation}
so that the dynamics of the scalar does not back-react on the evolution of the Universe at early times.

At this point, it is worth pausing to note that, in the present incarnation, the number of baryons produced depends on the dark matter density at some point in the past, but the necessity of tuning the other parameters in the model means that this model is not an example of asymmetric dark matter \cite{Zurek:2013wia}. However, the model presented here is a simple toy model for illustrative purposes and the implementation of our mechanism into more complicated models may allow for a direct connection between the densities of dark matter and baryons~\cite{Enqvist:1998en,Fujii:2002aj}.

\begin{figure}[h]
\centering
\stackunder{
\includegraphics[width=0.45\textwidth,height=0.45\textwidth]{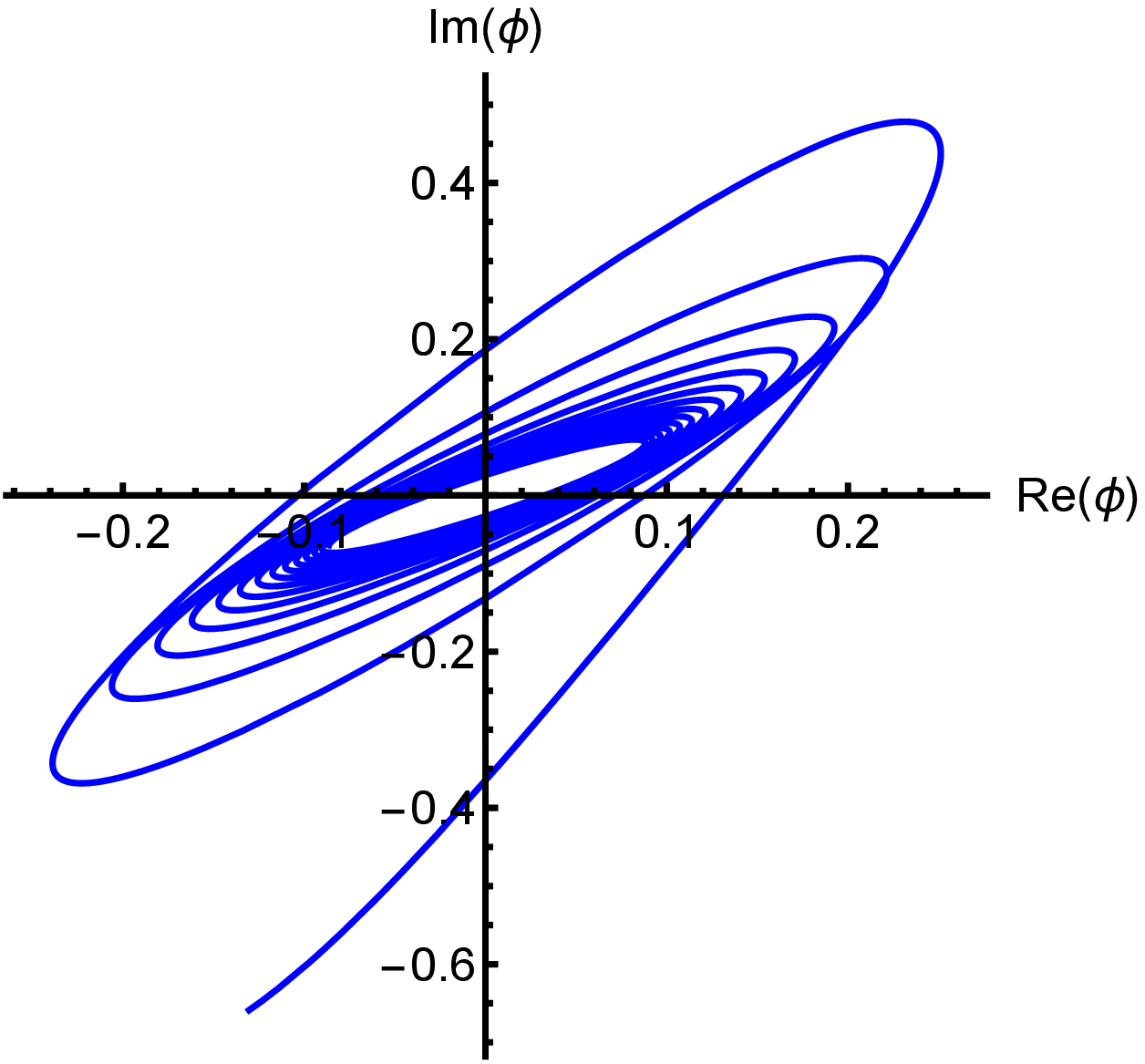}}{$n=2$}
\stackunder{
\includegraphics[width=0.45\textwidth,height=0.45\textwidth]{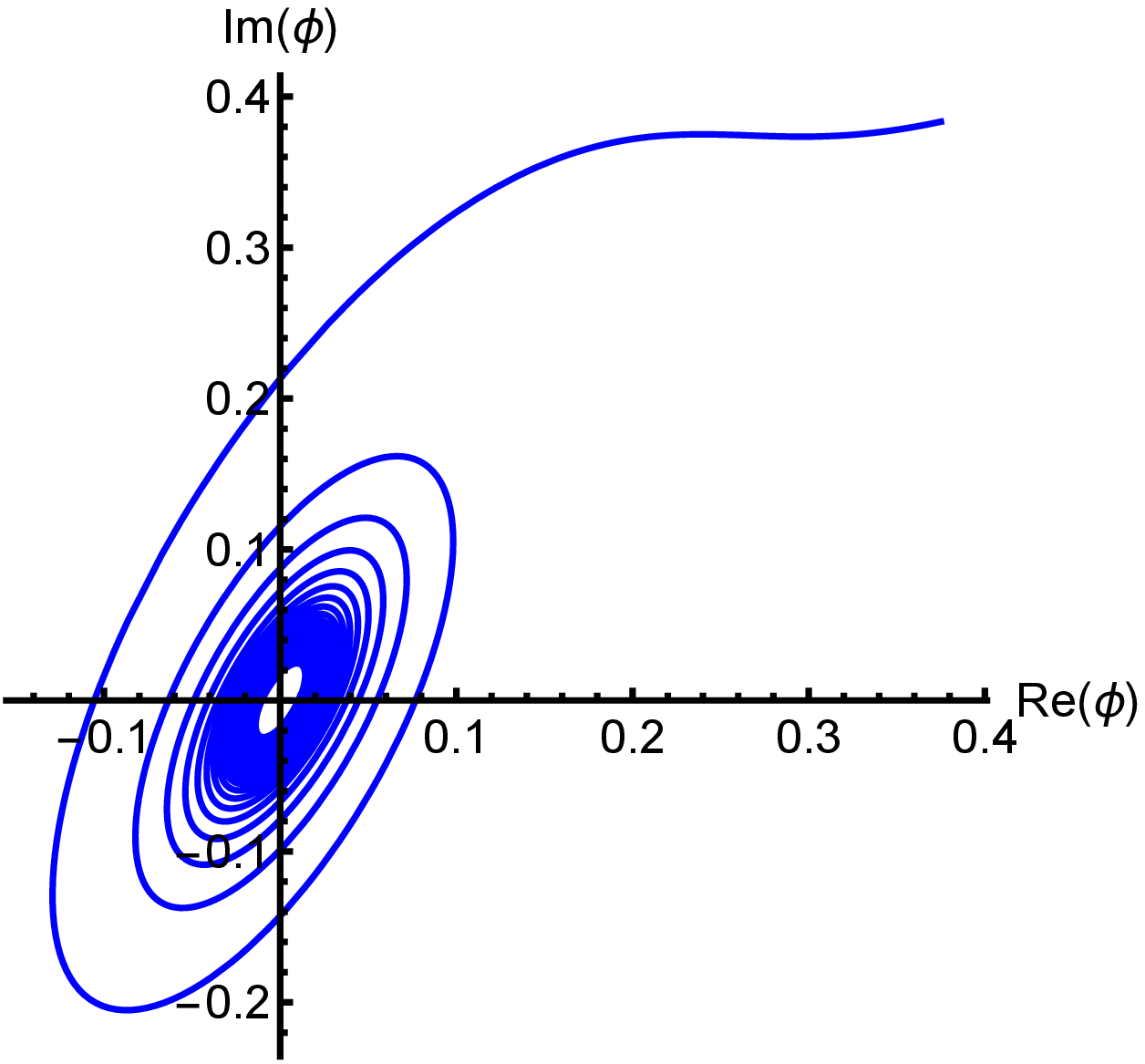}}{$n=3$}
\caption{The motion of the field for the cases $n=2$ and $n=3$, with parameters $\lambda=10^{-20}$ ($n=2$), $\mu=0.1 \mpl$ ($n=3$), $\varepsilon_0=0.03,\, \thee=3,\, \alpha=10^{10}$, and $m_\phi=0.1$ GeV. }
\label{fig:field}
\end{figure}

\begin{figure}[t]
\includegraphics[width=0.35\textwidth]{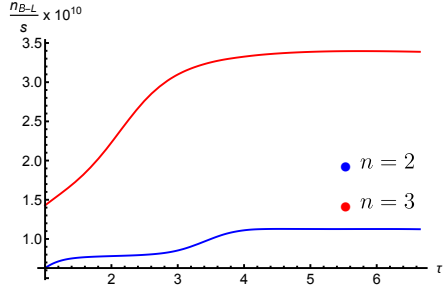}
\caption{The ratio $\nbl/s$ for the cases $n=2$ and $n=3$; the dimensionless time $\tau=m_\phi t$.}
\label{fig:nbl}
\end{figure}

\begin{figure}[t]
{\includegraphics[width=0.35\textwidth]{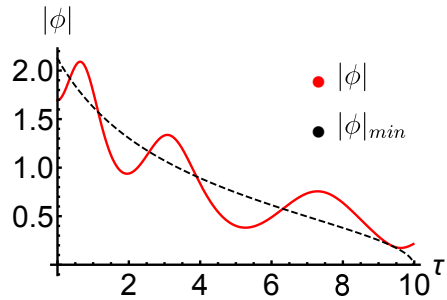}}
\caption{The motion of the absolute value of the field as a function of dimensionless time $\tau=m_\phi t$ for the case $n=2$. The red solid line shows the result of the numerical integration and the black dashed line shows the time-dependent minimum. The evolution is plotted up until $H=H\osc$.}
\label{fig:n2attractor}
\end{figure}

As mentioned earlier, we have made several approximations to arrive at our results and so we have numerically integrated the equations of motion to verify that our estimates are correct. As an example, in figure \ref{fig:field} we plot the evolution of the field for the cases $n=2$ and $n=3$ and one can see the expected behavior: the field tracks its minimum until the symmetry is restored at which point the angular field begins to rotate, generating a net $B-L$. The parameters used were $\lambda=10^{-20}$ ($n=2$), $\mu=0.1 \mpl$ ($n=3$), $\varepsilon_0=0.03,\, \thee=3,\, \alpha=10^{10}$, and $m_\phi=0.1$ GeV; we started the evolution from a universe with $\Omega_X=10^{-18}$ with $T_R=10^8$ GeV. These parameters were chosen to give $\nbl/s\sim10^{-10}$ according to equation \eqref{eq:nblfinal}. We plot this ratio in figure \ref{fig:nbl}; one can see that the numerically calculated value agrees with our prediction, validating the approximations we have employed. We also plot the motion of the field as a function of time in figure \ref{fig:n2attractor} for the case $n=2$ in order to verify our earlier assertion that the symmetry-breaking minimum is an attractor for the renormalizable theory, in contrast to the the case of Affleck-Dine baryogenesis during a phase of perturbative reheating. 

\clearpage

After the $B-L$ has been generated it is stored in the scalar condensate (it cannot decay through the symmetry-violating operators because $\phi$ is close to zero at this point and these are less important than the symmetry-preserving operators \cite{Lozanov:2014zfa}), which either decays, or, more likely, fractures into Q-balls~\cite{Kusenko:1997si,Kusenko:1997zq,Hertzberg:2013jba,Lozanov:2014zfa}. Q-balls are non-topological solitons that exist in any $\mathrm{U}(1)$ theory where the minimum of $V(|\phi|)/|\phi|^2$ occurs at some value other than $\phi=0$ \cite{Coleman:1985ki}, and are the end state of many Affleck-Dine models\footnote{The specific model we have studied here does not satisfy the condition for the existence of Q-balls (although this conclusion is likely to change once radiative corrections are accounted for) and so we expect the field to decay perturbatively. We expect that any more realistic generalization will likely end up as Q-balls. It would be interesting to investigate the properties of Q-balls that are coupled to dark matter in future work.}. They are unstable to decay to fermions \cite{Cohen:1986ct} and are therefore one way of transferring the asymmetry to the visible sector. We discuss how this may occur presently. 

\section{Mediation to the Visible Sector}\label{sec:SM} 
The Affleck-Dine mechanism naturally includes a method of transferring the asymmetry to the visible sector because $\phi$ typically corresponds to some flat direction for a combination of squarks or sleptons. In our scenario, we need to ensure that the asymmetry transfer is possible whilst still satisfying particle physics bounds. Here, we will focus on couplings to the standard model, although the implementation of the mechanism into beyond the standard model theories is certainly possible \cite{Brax:2006dc,Brax:2006kg,Hinterbichler:2010wu,Brax:2012mq,Brax:2013yja}. We will focus on one simple possible transfer mechanism as a proof of principle; there are numerous possible ways of incorporating a $\mathrm{U}(1)_{B-L}$ scalar into different particle physics models, even more so given that the nature of $X$ is not fixed by our mechanism, nor is the particle that comprises the dominant component of dark matter today. 

The lowest dimension operator that is a gauge singlet under the standard model gauge group is the neutrino portal operator $\bar{L}\tilde{H}$ \cite{Falkowski:2009yz,Escudero:2016tzx} (see \cite{Gonzalez-Macias:2016vxy} for a UV completion), where $L=(l \,\,\nu_l)^T$ is a left-handed lepton $\mathrm{SU}(2)$ doublet\footnote{In general, one has a Yukawa matrix for all three generations. We do not consider this here for simplicity.} and $\tilde{H}=i\sigma^2H^*$ is the charge-conjugated Higgs doublet. This operator only carries lepton number and so the simplest operator we can write down that couples it to the scalar is the dimension-five operator
\begin{equation}\label{eq:neuportal}
\frac{\phi\bar{L}\tilde{H}Y}{\Lambda} +\textrm{ h.c.}= g\phi\bar{\nu}_lY +\textrm{ h.c.},
\end{equation}
where $Y$ is a singlet fermion, $\Lambda$ is the cut-off for the effective field theory, $g=v/\sqrt{2}\Lambda$ with $v\sim 246$ GeV being the Higgs VEV, and the second equality holds in the broken-symmetry phase. Note that it is necessary to assign $\phi$ a lepton number $q=1$. In principle, one could take $Y$ to be $X$, but, as discussed above, there is no need to take $X$ to be a fermion, or, indeed, to be the dominant component of dark matter today, and so we prefer to keep our discussion as general as possible, especially since the operator \eqref{eq:neuportal} is by no means the unique coupling to the visible sector.

We require the scalar to decay to neutrinos in order to transfer the asymmetry and so we must impose $m_\phi>m_Y$. The phenomenology of this scenario has been investigated in \cite{Macias:2015cna}, in which an effective field theory of the neutrino portal was constructed\footnote{Note that the scalar considered there is real and not charged under any symmetry. For this reason, we have fewer operators than listed in \cite{Macias:2015cna}.}. Away from resonances when $m_\phi= m_Z,\, m_H$, the dominant self-annihilation process governing the relic abundance is $YY\rightarrow\bar{\nu}\nu$ mediated by the scalar, and one finds that the presently observed relic abundance for a cutoff given by
\begin{equation}
\Lambda \approx \left(\frac{70\textrm{ GeV}}{m_Y}\right)^{1/2} \left(1+\frac{m_\phi^2}{m_Y^2}\right)^{-1/2} \ , 
\end{equation}
where $m_Y$ is the mass of the fermion. 

Apart from this relatively weak constraint, the phenomenology of the neutrino portal is consistent with all other beyond the standard model searches \cite{Macias:2015cna}. The most important operators in the broken symmetry phase, besides the neutrino portal, are $vm_Z\bar{Y}\gamma^\mu Z_\mu Y/\Lambda^2$ and $vh\bar{Y}Y/\Lambda$, where $H=(v+h)/\sqrt{2}(0\;1)^{\rm T}$. The collider bounds on the invisible decay of the Higgs and the $Z$ are easily satisfied for these couplings, as are bounds from both spin-dependent and spin-independent nucleon scattering experiments. Indirect detection signals using dark matter self-annihilation in the Sun and the galactic halo are many orders of magnitude below current experimental bounds. 

\section{Equivalence Principle Violations} 
From the point of view of gravitational physics, our model is a scalar-tensor theory with dark matter moving on geodesics of the \emph{Jordan frame metric} $\tilde{g}_{\mu\nu}$ given by \eqref{eq:Weyl}. In the Newtonian limit, the effective metric for dark matter is 
\begin{equation}
\tilde{g}_{00}\approx \left(-1+2\Phi_{\rm N}+\frac{2|\phi|^2}{3M^2}\right) \ ,
\end{equation}
where $\pn$ is the Newtonian potential ($GM/r$ in the case of spherical symmetry), so that the effective Newtonian potential is $\tilde{\Phi}_{\rm N}=\Phi_{\rm N}+|\phi|^2/3M^2$. Dark matter therefore feels an additional or \emph{fifth}-force given by~\cite{Sakstein:2015oqa,Burrage:2016bwy}
\begin{equation}
F_5=\frac{\phi^\dagger\overset{\leftrightarrow}{\nabla}\phi}{3M^2} \ .
\end{equation} 
Baryons (and other forms of dark matter) do not feel this force and therefore fall at a different rate to $X$ particles in an external field, signaling a breakdown of the equivalence principle. If $X$ comprises the dark matter we observe today then, in theory, this could be tested by looking for equivalence principle violations between dark matter halos and their constituent galaxies \cite{Kesden:2006zb,Kesden:2006vz,Jain:2011ji}. The present day cosmological field has $\phi_0=0$, and therefore compact objects will only source a scalar field if they are dense enough to break the symmetry themselves (rather than due to the cosmological density) so that the field moves away from zero and towards its new minimum. By design, our model only exhibits symmetry breaking at densities well above the scale of the electroweak phase transition. Densities that high are certainly not reached in the cores of dark matter halos. In practice then, equivalence principle violations are expected to be negligible. A similar statement holds for cosmological probes such as the cosmic microwave background; the fifth-force leads to an enhanced growth of dark matter fluctuations, but only when the symmetry is broken \cite{Brax:2011pk}, i.e. well above big bang nucleosynthesis. 

\section{Baryon-Symmetric Baryogenesis}

Thus far, we have primarily focussed on an Affleck-Dine type scenario but this is by no means the only baryogenesis mechanism that generates a net $B-L$ using inverse symmetry-breaking transitions. Before concluding, we pause briefly to sketch how our mechanism can be implemented into one of these alternative scenarios: baryon-symmetric baryogenesis~\cite{Dodelson:1989ii,Dodelson:1989cq,Dodelson:1990ge,Dodelson:1991iv}.  

In this model, $\mathrm{U}(1)_B$ is a fundamental symmetry of the Lagrangian that is broken spontaneously at high temperatures when a hidden sector scalar acquires a time-dependent VEV. Baryon number is therefore not conserved separately in each sector so that one has $\partial_\mu J^\mu_{\rm hidden}=-\partial_\mu J^\mu_{\rm visible}$. A large baryon-anti-baryon asymmetry can then be generated in the visible sector that is exactly compensated by the charge stored in the hidden sector scalar condensate. The symmetry breaking requires two scalar fields $\phi$ and $\chi$, with $\phi$ charged under baryon number so that it can couple to right-handed quarks. The scalar potential is
\begin{equation}\label{eq:Vtreebsb}
V(\phi,\chi)=m_\phi^2|\phi|^2+\alpha_1|\phi|^4+\alpha_2|\chi|^4-2\alpha_3|\phi|^2|\chi|^2.
\end{equation}
Thermal corrections result in a symmetry breaking minimum at high temperatures for $\phi$ when $2\alpha_2>\alpha_3>2\alpha_1$ so that it acquires the necessary VEV for baryogenesis. An alternative to this potential, and the second field $\chi$, is to use our mechanism. This may open up the parameter space and give a richer phenomenology\footnote{One may also worry that a zero mass for $\chi$ is not technically natural since it is coupled to $\phi$, which in turn couples to the visible sector, and therefore a large mass for $\chi$ may be generated by quantum corrections. Our model does not suffer this problem since the field $\chi$ is absent.}.

\section{Summary and Conclusions} 
In this paper we have identified a new baryogenesis mechanism based on the novel breaking of a $\mathrm{U}(1)$ symmetry in the early Universe, in which the Weyl coupling of a charged scalar to dark matter induces a tachyonic mass at sufficiently high densities. The symmetry is ultimately restored when the energy density in dark matter redshifts to a small enough value. Many existing ideas for baryogenesis rely on inverse-symmetry breaking transitions and here we have given an example of how our mechanism can be implemented in a straightforward manner into one common and well-studied paradigm: the Affleck-Dine mechanism. We have illustrated the new features and phenomenology that arise due to the dark matter coupling, and have demonstrated that a fully consistent scenario can be readily obtained. In particular, the asymmetry generated can be transferred to the visible sector without violating current experimental bounds on new particles coupled to the standard model. In this work we have presented a simplified toy model for clarity. In future work we will investigate whether this scenario might find a natural home in models of physics beyond the standard model. 

\section*{Acknowledgements} We are grateful for useful discussions with Mustafa Amin, Clare Burrage, and Peter Millington. The work of J.S. was supported by funds provided to the Center for Particle Cosmology by the University of Pennsylvania. The work of M.T was supported in part by US Department of Energy (HEP) Award DE-SC0013528, and also by NASA ATP grant NNX11AI95G.

\bibliography{ref}

\end{document}